# The Composition and Mineralogy of Rocky Exoplanets: A Survey of >4,000 Stars from the Hypatia Catalog

Keith D. Putirka[1,2]
John C. Rarick[1,3]

[1]Department of Earth and Environmental Sciences, Fresno State, 2345 E. San Ramon Ave, MS/MH24, Fresno, CA 93720, 559-278-4524; [2]kputirka@csufresno.edu; [3]jcrarick@outlook.com

## Abstract

We present a survey of >4,000 star compositions from the Hypatia Catalog to examine whether rocky exoplanets (i.e., those with rocky surfaces, dominated by silicates) might be geologically similar to Earth, at least with respect to composition and mineralogy. To do so, we explore the variety of reported stellar compositions to then determine a possible range of exoplanetary mantle mineralogies. We find that exoplanetary mantles will likely be dominated by olivine and/or orthopyroxene, depending upon Fe partitioning during core formation. Some exoplanets may be magnesiowüstite- or quartz-saturated, and we present a new classification scheme based on the weight % ratio (FeO+MgO)/SiO$_2$, to differentiate rock types. But wholly exotic mineralogies should be rare to absent. We find that half or more of the range of exoplanet mantle mineralogy is controlled by core formation, which we model using $\alpha_{Fe} = Fe^{BSP}/Fe^{BP}$, where $Fe^{BSP}$ is Fe in a Bulk Silicate Planet (bulk planet, minus core), on a cation weight % basis (elemental weight proportions, absent anions) and $Fe^{BP}$ is the cation weight % of Fe for a Bulk Planet. In our solar system, $\alpha_{Fe}$ varies from 0 (Mercury) to about 0.54 (Mars)]. Remaining variations in exoplanet mantle mineralogy result from non-trivial variations in star compositions. But we also find that Earth is decidedly non-solar (non-chondritic); this is not a new result, but appears worth re-emphasizing, given that current discussions often still use carbonaceous or enstatite chondrites as models of bulk Earth. We conclude that such models are untenable, regardless of the close overlap of some isotope ratios between certain meteoritic and terrestrial (Earth-derived) samples. There is also the possibility that Earth contains a hidden component, that if added to known reservoirs would yield a solar/chondritic Earth. We test that idea using a mass balance of major oxides using known reservoirs, so that the sum of upper mantle, metallic core and crust, plus a hidden component, yield a solar bulk composition. Here, the fractions of crust and core are fixed and the hidden mantle component is some unknown fraction of the entire mantle, $F_h$ (so if $F_{DM}$ is the fraction of depleted mantle, then $F_h + F_{DM} = 1$). Such mass balance shows that if a hidden mantle component were to exist, it must comprise >28% of Earth's mantle, otherwise it would have negative major oxide abundances. There is no clear upper limit for such a component, so it could comprise the entire mantle, but all estimates from $F_h = 0.28$ to $F_h = 1.0$ yield a hidden fraction that does not match the sources of mantle plumes or mid-ocean ridge basalt (MORB). The putative component is also geologically unusual, being enriched in Na and Fe, and depleted in Ca and Mg, compared to familiar mantle components. We conclude that such a hidden component does not exist.

## Introduction

Rapid and numerous discoveries of exoplanets have emerged from the Kepler mission (e.g., Thompson et al. 2018), which relies on the dimming of light from an observed star, when a planet passes within line of sight, providing a partial stellar eclipse. The mission has shown that exoplanets are common if not ubiquitous. That recognition has induced more than a little curiosity about whether rocky exoplanets might exhibit plate tectonics (e.g., Weller and Lenardic 2018), and how geologic processes might be connected to the evolution of atmospheres, oceans, and life (e.g., Stern 2006; Foley and Driscoll 2017). But this curiosity has not been matched by knowledge of potential exoplanet compositional diversity. Are any exoplanets utterly exotic, made of mostly oxides, or a strange mix of silicates? Or are they mostly like Earth? And do interstellar composition variations or intra-planetary system processes exert hegemony over planetary compositions? (We use "terrestrial" to mean rocks that are from or "of Earth"; for exoplanets that are like Earth, or other planets in our inner solar system, we use "rocky" as an adjective). To answer these questions we explore a range of star compositions so as to determine a likely range of exoplanetary mantle mineralogies. We do so using the Hypatia Catalog (Hinkel et al. 2014), which provides highly precise compositions of nearby stars, a fraction of which are the targets of the Kepler mission. A foundation of such exoplanet research (e.g., Tachinami et al. 2011; Duffy et al. 2015; Unterborn et al. 2017) is the long-observed similarities of non-volatile element abundances between chondrite meteorites and the solar photosphere (Pottasch 1964; Ringwood 1966; Lodders and Fegley 2018); their close match implies that planets should be similar in composition to the stars they orbit. We also test this assumption.

Naturally, a key question is what fraction of exoplanets might be Earth-sized and or have a similar Earth-Sun distance. The Kepler discoveries are observationally biased towards detection of planets with brief orbital periods (which then provide multiple observations in a short time period) and/or large planets that orbit small stars (providing a greater signal of dimming), as is also the case for the most recent discoveries from the TESS mission (Vanderspek et al. 2018). But the new discoveries show that exoplanets are not oddities. Dressing and Charbonneau (2013), for example, show that as many as 51% of exoplanets that orbit small stars range from 0.5—1.4 times Earth's radius, while a new statistical model of Sun-like Stars by Mulders et al. (2018) indicates an occurrence rate among stellar systems of up to 100% for exoplanets with 1AU-sized orbits. Within our galaxy, Earth-sized planets with near-1AU orbits are thus likely quite common.

Although knowledge of extrasolar planet compositions is scanty, some recent studies have begun to fill the gap, using the assumption that rocky exoplanets have similar compositions to the stars they orbit, and applying Gibbs Free Energy Minimization models (GFEMs) to predict silicate mantle mineralogies for about a dozen planets (Unterborn et al. 2017; Hinkel and Unterborn 2018). But GFEMs are time-intensive and so not readily employed to survey large numbers of exoplanet compositions. And no GFEM has been tested to predict mineral proportions in natural samples (e.g., see tests of our approach in Fig. 1, discussed later). Moreover, to anticipate other of our conclusions, current experimental data do not span the range of MgO-poor, $SiO_2$-rich compositions (<20 wt. % and >50 wt. % respectively; Fig. 2b) observed for some exoplanets, and so untested GFEMs (and our models) require extrapolation.

As an alternative we propose a two-fold approach, which allows an evaluation of a much greater fraction of stellar systems in the Hypatia Catalog, yields a clearer analysis of error, and reveals voids in current data sets that must be filled. We begin by comparing rocky exoplanet compositions to terrestrial rocks and minerals; from these comparisons, we derive a classification scheme that yields exoplanetary mantle rock types from weight % ratios of (FeO+MgO)/$SiO_2$. We also use Thompson's (1982) algebraic mass balance approach to recast major oxides into mineral proportions, with mineral compositions derived from experimentally-equilibrated systems. As Thompson (1982) noted, this approach can "save you time and money"—and so it does, allowing us to estimate mineral proportions for thousands of exoplanet compositions, and to compare these to terrestrial compositions using the ultramafic rock diagram of Le Bas and Streckeisen (1991).

Using this approach we examine >4,000 stars, or >80% of the Hypatia Catalog (Hinkel et al. 2014). Our approach allows us to assess how assumptions of planetary temperature affect mineral proportions, which motivates our proposal of a "standard mineralogy", as we discuss in the Methods section. We also test the effects of core formation, taking Mercury, Earth and Mars as examples of mantle/bulk-planet Fe partitioning, which we find to have a substantial effect on calculated mantle mineralogies.

## Methods
### General

We estimate the range of exoplanet compositions and mineralogies using 4,382 stars from the Hypatia Catalog, taking all those entries where each of Na, Mg, Al, Si, Ca, Ti, Cr, Fe, and Ni are reported. We choose the Hypatia catalog for several reasons. First, it is the largest catalog that examines nearby (within 150 pc) F-, G-, K-, and M-class (main-sequence, or H-burning) stars; the Sun, for reference, is G-class (specifically G2V) and the Milky Way is 26.8 kpc in diameter (Goodwin et al. 1998). Second, by focusing on nearby stars, Hinkel et al.'s (2014) catalog has data of higher precision, for a broader range of elements. Many stars only have reports of, say, Fe/H and/or Si/H; we require an array of precise elemental analyses to obtain more realistic estimates of exoplanet mineralogy, and to test for the occurrence rate of mineralogical oddities. Third, Hinkel et al. (2014) provide errors, or the "spread" of star compositions, which they obtain by comparing multiple composition reports of a given star (see Hinkel et al. 2016). This is important as we wish to propagate such uncertainties into errors on exoplanet mineralogy. A final advantage is that the Hypatia catalog, by focusing on nearby stars, is highly relevant to the Kepler and TESS exoplanet-hunting missions.

To obtain a Bulk Silicate Planet (BSP) composition (the bulk planet, minus its metallic core), we assume that exoplanets have non-volatile element abundances that match the stars they orbit. This assumption, fundamental to exoplanet studies (e.g., Tachinami et al. 2011; Duffy et al. 2015; Unterborn and Panero 2017; Unterborn et al. 2017; Hinkel et al. 2018), stems from a remarkable 1-to-1 correlation of non-volatile element abundances between the Sun's photosphere and CI chondrite meteorites (e.g., Pottasch 1964; Ringwood 1966; Lodders and Fegley 2018), and appears validated by a GFEM-approach of nebular condensation (Thiabaud et al. 2015a). But as we will show, this assumption is imperfect.

Once we have a bulk planet (BP) composition, the silicate portion, BSP, is obtained by subtracting a metallic core. To form a metallic core, we use the ratio $\alpha_{Fe}$ = $Fe^{BSP}/Fe^{BP}$, where $Fe^{BSP}$ is Fe in a BSP, on a cation weight % basis (elemental weight proportions, absent anions) and $Fe^{BP}$ is the cation weight % of Fe for the Bulk Planet. Each of the planets Mercury, Earth and Mars have proportionately different sized cores, and non-overlapping $\alpha_{Fe}$, which we calculate assuming either that $Fe^{BP} = Fe^{BSP} + Fe^{Core}$, or that $Fe^{BP} = Fe^{Solar}$, where $Fe^{Solar}$ is also the cation fraction of Fe, here on a non-volatile basis, and similar to the Fe cation fraction of carbonaceous chondrites. As we detail below, we use Mercury, Earth and Mars as case studies: $\alpha_{Fe}^{Mercury}$ = 0.0-0.12; $\alpha_{Fe}^{Earth}$ = 0.263-0.494; $\alpha_{Fe}^{Mars}$ = 0.54-0.58. For Earth, we assume that the metallic core is 33% by mass and has 5 wt. % Ni (Rubie et al. 2011), 2 wt. % S (Rubie et al. 2011; Wood et al. 2014) and 4-7 wt. % Si (Wade and Wood 2005), leaving a core with 86-89 wt. % Fe. We use 87 wt. % Fe in the core for all subsequent calculations. Rubie et al. (2011) also estimate that Earth's core contains 0.5 wt. % O. But rather than model the uncertain behavior of O during core formation, we instead employ $\alpha_{Fe}$, as a combined proxy for $fO_2$ and core formation efficiency. As to BSP compositions, we also assume that whatever little C and S are retained following accretion is segregated to near surface environments rather than retained in the mantle (Fig. 2d). We test the fundamental assumption that stars provide an unfiltered view of exoplanet compositions using the Sun (Lodders 2010) as input, to obtain a Solar Bulk Silicate Planet composition, referred to as Sol-BSP. In principle, for some value of $\alpha_{Fe}$, Sol-BSP should match estimates of Bulk Silicate Earth (BSE; McDonough and Sun 1995; Salters and Stracke 2003; Workman and Hart 2005; Palme and O'Neill 2014).

## Stellar Oxygen Abundances and Formation of a Metallic Core

All Fe that is not partitioned into the core is treated as FeO total (FeOt). A critical question, is whether exoplanets have sufficient O to oxidize all cations of interest, let alone Fe. Unterborn and Panero (2017) find that if Earth is solar, O would be sufficiently abundant to oxidize all of Earth's core. We obtain a similar result for exoplanets: of 3,266 stars in the Hypatia Catalog where Si, Mg, Fe, O and C are all reported, 97.8% have sufficient O to oxidize not just all of Si, Mg and Fe (to $SiO_2$, MgO and FeO), but all available C to form $CO_2$. This is not to say that O (or C) abundances are not important, but we recognize that mantle C-O abundances are affected by a myriad of processes, including nebular condensation (e.g., Thiabaud et al. 2015b; Unterborn and Panero 2017), accretion (Rubie et al. 2015; Schaefer and Fegley 2017), and post-accretion outgassing (Wade and Wood 2005; Schaefer and Fegley 2010); combining these to ascertain a net planetary $fO_2$ is a fraught endeavor. Rather than estimate $fO_2$, our use of $\alpha_{Fe}$, as described above, serves as a combined proxy for $fO_2$ and core-formation efficiency, and obviates the need for employing highly uncertain models.

Our use of $\alpha_{Fe}$ is specifically meant to examine how different scenarios of core formation affect silicate mantle mineralogy. As in Unterborn et al. (2017), a fraction of bulk planetary Fe (and Ni and Si) is removed to form the core, with the remainder staying in the mantle. This process necessarily provisions the residual mantle with greater amounts of Si, Mg, and other cations which, as we show, greatly impacts mantle mineralogy. As noted, for the ratio $\alpha_{Fe} = Fe^{BSP}/Fe^{BP}$, the value of $Fe^{BSP}$ is the cation weight % of Fe in a Bulk Silicate Planet (BSP), relative to $Fe^{BP}$, the cation weight % of Fe for the bulk planet, core included, so $Fe^{BP} = Fe^{BSP} + Fe^{Core}$. For Earth, we examine three cases. In two of these, we accept that total Fe in the mantle, as FeO (FeOt, where "t" is "total") is 8 wt. % (McDonough and Sun 1995; Salters and Stracke 2003; Workman and Hart 2005; Palme and O'Neill 2014), which translates to $Fe^{BSP} = 11.3$ wt. %, on a cation basis. If the bulk Earth has a solar bulk composition, $Fe^{BP} = 42.9$ wt. %, and $\alpha_{Fe}^{Earth}$ is 0.263. If instead, we add to the mantle a core that is 33% Earth's mass and is 87 wt. % Fe, we have $Fe^{BP} = 36.61\%$ and $\alpha_{Fe}^{Earth} = 0.311$. As a third case we assume that Earth's bulk mantle FeOt is unknown, and subtract Earth's core (87 wt. % Fe, 33% Earth's mass) from a solar bulk composition, which yields $\alpha_{Fe}^{Earth} = 0.494$, and a mantle with 21.2 wt. % Fe (so about double the estimates for BSE).

For the cases of Mars and Mercury, a solar bulk composition requires $\alpha_{Fe}^{Mars} = 0.537$ and $\alpha_{Fe}^{Mercury} = 0.116$ to respectively obtain a martian mantle with 17.3 wt. % FeOt (Taylor 2013) and a mercurian mantle with 3.5 wt. % FeOt (Morgan and Anders 1980). If we instead add a pure Fe core (Rubie et al. 2015) of 22% planetary mass for Mars, and 68% planetary mass for Mercury, to silicate mantles with 17.3% FeOt for Mars and 3.5wt. % FeOt for Mercury, we obtain $\alpha_{Fe}^{Mars} = 0.58$ and $\alpha_{Fe}^{Mercury} = 0.07$. But $\alpha_{Fe}^{Mercury}$ may be close to zero. Charlier et al. (2013) imply a nearly FeOt-free mercurian mantle. Additionally, Keil (2010) suggests that the parent body of enstatite achondrite (EA) meteorites is nearly Fe-free, and that the parent body might be Mercury. So we use $\alpha_{Fe}^{Merc/EA} = 0$ as a possible end-member case for our solar system.

To complete the model of the core-formation process we add Ni and Si into the metallic cores of exoplanets by applying an Earth-like $\alpha_{Ni} = Ni^{BSP}/Ni^{BP} = 0.08$ (where $Ni^{BSP} = 0.196$ wt. %, from McDonough and Sun, 1995; $Ni^{BP} = Ni^{Solar} = 2.54$ wt. %), and $Si^{BSP} = [1 - C_{Si}^{Core} X^{core}]C_{Si}^{Stellar} = [1 - (0.07)(0.33)]C_{Si}^{Stellar} = [0.98]C_{Si}^{Stellar}$. Of course, light-alloying element abundances in a metallic core might vary with planet size (Wade and Wood 2005), but our results show that dissolving more or less Si (or Ni) into the core makes very little difference to bulk mantle mineralogy.

## A Standard (Normative) Mineralogy & The Effects of Planetary Temperature

As noted in the Introduction, we use the algebraic methods of Thompson (1982) to recast oxide compositions into mineral proportions, which requires a choice of mineral compositions. This choice should be viewed as a "standard", or normative mineralogy, as these terms were used for terrestrial igneous rocks by Cross et al. (1902). A normative approach is useful here for the very reasons that it was useful in the early 1900s. Calculated mineral norms were never viewed as estimates of actual mineralogy, but instead allowed for discussions of compositional contrasts in mineralogical terms. At that time, geologists knew quite little about the pressure-temperature (*P-T*) conditions of igneous crystallization, or crystallization rates (*dT/dt*), but they suspected that mineralogy depended upon these. Cross et al. (1902) thus proposed to calculate the proportions of end-member mineral compositions, which they called "standard" or "normative" minerals, which stood in contrast to the natural minerals, whose relative abundances were called "modes".

For exoplanets, the problems are the same: exoplanetary $T$-$fO_2$ conditions (and *P*, for planets of varying size) are effectively completely unknown, and $T$ and $fO_2$ vary with time in any case, as planets cool and differentiate. Therefore, we estimate the proportions of minerals using a "standard" mineralogy (see Fig. 3 caption), using mineral compositions from the experiments of Walter (1998) and sub-solidus compositions used in Putirka (2008); these compositions were selected so as to reproduce the bulk mineralogy of depleted mantle (DM) of Workman and Hart (2005) when the Workman and Hart (2005) DM bulk oxide composition is used as input. Our standard mineralogy approximates equilibrium at 1350°C at 2.0 GPa, but can be found in experiments equilibrated from 1225-1450°C and 1-2.5 GPa. Our standard minerals thus also imply a standard set of *P-T* conditions. These conditions are low enough to apply to any planet whose mantle reaches pressures of at least 2 GPa, which will include any differentiated planet larger than Earth's Moon (whose core/mantle boundary is at 5 GPa; for reference, Mercury's mantle reaches pressures of 8 GPa; Antonangeli et al. 2015). We use such compositions, as opposed to idealized end-members, in the hope that our calculated mineral proportions (Fig. 3) might approximate actual exoplanet mineralogies, and Fig. 1 provides some grounds for that aspiration: calculated fractions of olivine (Ol), orthopyroxene (Opx) and clinopyroxene (Cpx) closely match measured mineral fractions in natural peridotites (Warren 2016) and pyroxenites (Bodinier et al. 2009), at least when Cpx fractions are < 0.4, (the case for all exoplanets we examine). Our tests imply that mineral modes are affected more by bulk composition than solid solution. But Opx, Ol and Cpx exhibit sufficient solid solution to affect plotted mineral proportions in a Le Bas and Streckeisen (1991) diagram (Fig. 3). In that figure we show arrows to indicate how compositions shift if one were to apply pure end-member compositions, which are a proxy for a low temperature case of 800-950°C. We also use the noted experiments to determine solid solution limits of Ol, Opx and Cpx, and so test whether exoplanets are sufficiently rich in Ti, Si, or Al to require saturation of phases such as rutile, quartz or corundum, etc.

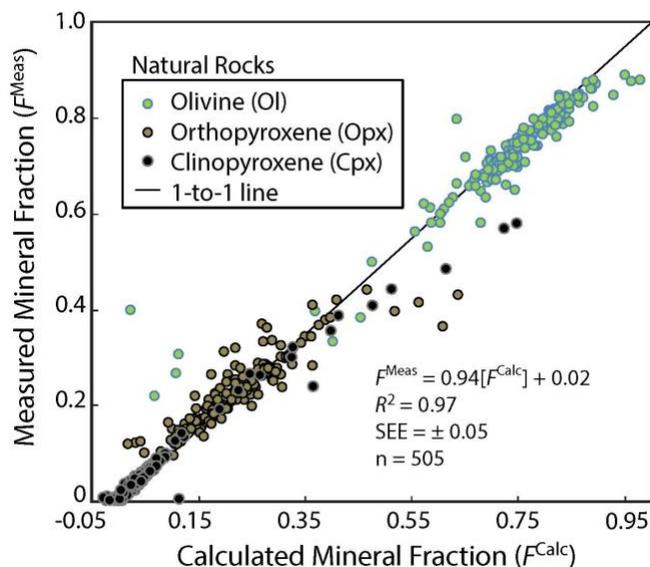

**Figure 1.** Mineral fractions calculated using Thompson (1982) ($MF^{Calc}$) are compared to 550 precisely-measured mineral fractions ($MF^{Meas}$) derived from 168 natural peridotite and pyroxenite bulk compositions (Bodinier et al. 2008; Warren 2016); 99% of these samples contain spinel and all calculated and measured mineral fractions are normalized so that Ol + Opx + Cpx = 1 (e.g., for triangular plotting, as in Fig. 2). Observed modes range from 0-89% Ol, 9-44% Opx and 0.3-58% Cpx. Inset equation gives results for the total regression on all 550 mineral abundance estimates. Uncertainties on individual mineral grains are similar, but note systematic error on calculated Cpx fractions at Cpx fractions > 0.4. Also note that observed Opx fractions and much less than expected for some exoplanets, which may approach nearly 100% Opx

Results from GFEMs should be quite close to ours, and we do not suggest that GFEMs are necessarily inaccurate; we only emphasize that they are an inefficient means to process thousands of samples, and have not been tested as in our Fig. 1. There should be no intrinsic bias to either method in that we expect the laws of thermodynamics to be the same, across time, and across the Milky Way. But our experimental data are limited. A significant fraction of exoplanet BSPs (EBSPs) have simultaneously low MgO (<20wt. %) and high $SiO_2$ (> 50 wt. %), and these compositions fall outside the range of current experimental studies. Experiments on these compositions may reveal new phases or even new solid solution limits, and provide GFEMs with better control on the saturation of phases such as quartz, rutile or corundum, etc..

## Calculation Procedure: Converting Stellar Dex System Compositions to Silicate minerals

An exoplanetary silicate mantle composition and mineralogy is obtained as follows:
(1) Star compositions from the Hypatia Catalog (Hinkel et al. 2014), in dex-system notation, are converted to weight % of elements (Supplementary Table 1) using the Solar Photosphere composition of Lodders (2003, 2010). Total stellar Fe (which we take as $Fe^{BP}$) is multiplied by $Fe^{BSP}/Fe^{BP}$, considering separate cases for Earth ($\alpha_{Fe}^{Earth}$ = 0.263-0.494), Mercury ($\alpha_{Fe}^{Mercury}$ = 0.0-0.12), and Mars ($\alpha_{Fe}^{Mars}$ = 0.54-0.58), to obtain estimates of $Fe^{BSP}$.

(2) Total Ni, or $Ni^{BP}$, is multiplied by $\alpha_{Ni}$ = 0.11, and total Si, or $Si^{BP}$, is multiplied by $\alpha_{Si}$ = 0.93 to obtain $Ni^{BSP}$ and $Si^{BSP}$ respectively (Supplementary Table 2).

(3) The entirety of Ti, Ca, Al, Cr, and Na are retained in the mantle, and we renormalize so that Ti + Ca + Al + Cr + Na + $Fe^{BSP}$ + $Ni^{BSP}$ + $Si^{BSP}$ = 1.

(4) Elemental abundances are converted to oxide weight %, with Fe calculated as FeOt.

(5) We apply the methods of Thompson (1982) to recast, for example, the oxides $SiO_2$, $Al_2O_3$, FmO (FmO = FeO + MgO), and CaO, as proportions of Olivine (Ol) = $Fm_2SiO_4$, Orthopyroxene (Opx) = $Fm_{1.9}Ca_{0.1}Al_{0.2}Si_{1.8}O_6$, Clinopyroxene (Cpx) = $Ca_{0.6}Fm_{1.4}Al_{0.2}Si_{1.8}O_6$ and Garnet = $Fm_{2.7}Ca_{0.3}Al_2Si_3O_{12}$, where FmO = FeO+MgO (Supplementary Table 2). We also calculate mineral proportions using pure minerals olivine ($Fm_2SiO_4$), clinopyroxene ($CaFmSi_2O_6$), orthopyroxene ($Fm_2Si_2O_6$), and garnet ($Fm_3Al_2Si_3O_{12}$).

As we will show, most, but not all, exoplanets can be described as a positive combination of Ol + Cpx + Opx + Gar. But we also examine minor phases, such as rutile ($TiO_2$), chromite ($FeCr_2O_4$), corundum ($Al_2O_3$), quartz ($SiO_2$), albite ($NaAlSi_3O_8$), nepheline ($NaAlSiO_4$) and bunsenite (NiO), to name a few, that might be globally saturated.

### Testing the Assumption that Stellar and Planetary Compositions are Identical

As a test of our Methods, we treat the Sun as it were just another star in the Hypatia Catalog, deriving a BSP with a solar bulk composition (Sol-BSP). This is then compared to estimates of Bulk Silicate Earth (BSE; McDonough and Sun 1995; Palme and O'Neill 2012; Table 1) and depleted MORB-source mantle, DM (Salters and Stracke 2003; Workman and Hart 2005) (published estimates of DM and BSE are effectively identical). We also compare Bulk Silicate Planet (BSP) estimates for Mars (see Taylor 2013) and Earth's Moon (see Kahn et al. 2006, and Longhi's 2006 LPUM) and we calculate a new BSE (Table 1) that accounts for a compositionally distinct Plume Source Mantle (PSM) for ocean island and flood basalts that appear to be a mixture of DM and subducted mid-ocean ridge basalt (MORB) (see Putirka et al. 2018 and references therein). For our new BSE, we assume that plumes tap some fraction, x, of Earth's lower mantle, so that BSE = xPSM + (x-1)DM; Table 1 illustrates the case of x = 0.73, where PSM constitutes the entirety of Earth's lower mantle.

## Results & Uncertainties

### Exotic Mineralogies are Rare to Absent

Among the >4,000 stars examined, we find that $SiO_2$, MgO, and FeO are the dominant oxides, comprising ≥ 80% of all oxides for all stars examined. Like Hinkel and Unterborn (2018), we find that the Sun is slightly enriched in Fe relative to the median of Hypatia stars (Fig. 2a) and also has values close to the Hypatia median for other oxides (e.g., MgO, $SiO_2$ and $Na_2O$; Fig. 2). But precise BSP compositions are sensitive to $\alpha_{Fe}$. For example EBSPs in Fig. 2a, for the case of $\alpha_{Fe}$ = 0.263, are shown as gray circles, whereas red circles show exoplanets when $\alpha_{Fe}$ = 0.311; the latter have higher FeO by about 2 wt. %. For the case of $\alpha_{Fe}$ = 0.263, the range of exoplanetary FeO in Fig. 2a implies silicate mantles that have 2-12 wt. % FeOt (median = 7.7 wt. %), and metallic cores of pure Fe would vary proportionately in radius from -32% to +22% relative to Earth ($r_{median}^{core}$ = 3,421 km; $r_{Earth}^{core}$ = 3,480 km).

Submitted to *American Mineralogist*

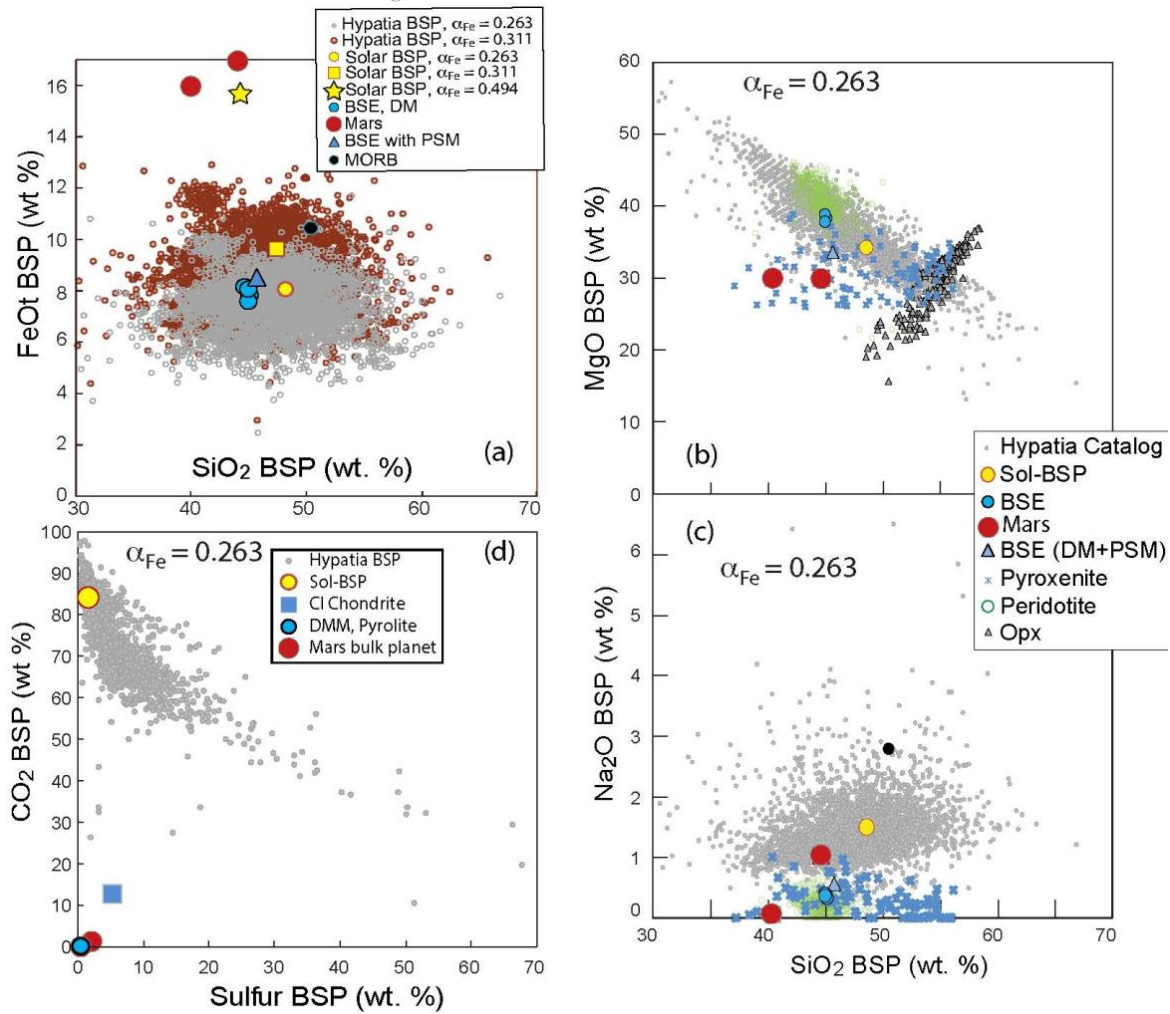

**Figure 2.** (a) FeOt vs. SiO$_2$ (weight %) for exoplanet Bulk Silicate Planet (BSP) compositions, calculated assuming two different models for the partitioning of Fe between core and mantle: $\alpha_{Fe}$= 0.263 (gray circles) and $\alpha_{Fe}$= 0.311 (red circles). These are compared to estimates of bulk silicate Earth (BSE), Earth's Depleted Mantle (DM), and a new BSE that assumes that Earth's entire lower mantle is Plume Source Mantle (BSE with PSM; blue triangle), which is a combination of 80% DM + 20% Mid-Ocean Ridge Basalt (MORB; black dot), and three examples of Sol-BSP derived assuming $\alpha_{Fe}$= 0.263, $\alpha_{Fe}$= 0.311 and $\alpha_{Fe}$= 0.494. See text for explanation and Table 1 for bulk compositions. Exoplanet bulk compositions, and calculated BSPs are provided in Electronic Appendix A. Also plotted are SiO$_2$ vs. MgO (b) and Na$_2$O (weight %) (c) for exoplanet BSPs (EBSP) when $\alpha_{Fe}$= 0.263. Panels (b) and (c) compare exoplanets to terrestrial peridotites (green circles; GEOROC; http://georoc.mpch-mainz.gwdg.de/georoc/) and pyroxenite bulk compositions (blue stars) from experimental studies in the LEPR database (Hirschmann et al. (2008). (b) also shows terrestrial orthopyroxene (Opx) compositions (LEPR); Opx has the highest SiO$_2$ of common ultramafic minerals, and so rocks to the high-SiO$_2$ side of Opx, e.g., with SiO$_2$> 0.56[MgO] – 40 are plausibly quartz-saturated. Exoplanets that overlap with the Opx field are possibly mono-mineralic. In (c) exoplanet BSPs likely have much less Na$_2$O than plotted since Na is volatile—which explains why Earth has as little as 25% of Sol-BSP of Na$_2$O (which presumes that all Na is retained). (d) shows CO$_2$ (wt. %) vs. S (weight %) for EBSPs where Fe, Mg, Si, C, O, and S are all reported, and where we assume that all C and S are retained during condensation and accretion; these are compared to Sol-BSP, bulk Mars, BSE, DM, and CI chondrites. Earth and Mars have clearly lost most of their C and S.

Figure 2a also illustrates how assumptions of $\alpha_{Fe}$ affect estimates of Sol-BSP, which represents the silicate mantle of a planet of Solar bulk composition: yellow symbols represent calculations for the cases $\alpha_{Fe}$ = 0.263, 0.311 and 0.494. None of the estimates of Sol-BSP intersect estimates of Bulk Silicate Earth (BSE; McDonough and Sun 1995; Palme and O'Neill 2012) or Depleted Mantle (DM, Salters and Stracke 2003; Workman and Hart 2005). Neither do the Sol-BSP estimates match our new "BSE with PSM"; this estimate is derived using the plume source composition from Putirka et al. (2011, 2018), where PSM = 20%

Mid-Ocean Ridge Basalt (MORB; Gale et al. 2013) plus 80% DM, and assumes that PSM encompasses all of the lower mantle (Fig. 2a). Even with such large amounts of PSM, this new estimate of BSE does not reach the trend-line formed by the yellow symbols of Sol-BSP from various values of $\alpha_{Fe}$. If we instead assume that PSM only comprises the bottom-most portions of the lower mantle, then "BSE with PSM" is driven away from MORB (black circle) and towards DM (blue circles), with the result that "BSE with PSM" would then be little different from the BSEs of McDonough and Sun (1995) and Palme and O'Neill (2012).

In Figure 2b, we compare exoplanets at $\alpha_{Fe} = 0.263$ with terrestrial peridotites (green circles) and pyroxenites (blue stars) (Fig. 2b; see caption for data sources). Exoplanets clearly exhibit a wide range of MgO and $SiO_2$. At the high-MgO end of the peridotite array (green circles, Fig. 2b) our terrestrial samples are dunites, and BSPs that range to greater MgO are probably magnesiowüstite (Msw)-saturated. At the low-MgO/high $SiO_2$ end of the array (Fig. 2b), exoplanets range to higher $SiO_2$ than pyroxenites (blue stars, Fig. 2b), and orthopyroxenes; since Opx has the highest Si content among common mantle phases, such exoplanets may be saturated with quartz (Qtz). Aside from Msw and Qtz, it is not clear that phases that are minor on Earth are dominant elsewhere. Clinopyroxene (Cpx) and garnet (Gar) appear to exhibit sufficient solid solution capacity so as to absorb the small amounts of exoplanetary Ca, Al and Ti, while olivine (Ol) can absorb Ni.

Sodium might be an exception, but the issue is unclear as Na is volatile (Palme 2000). In Figure 2c, $Na_2O$ in exoplanet BSPs are maxima, calculated assuming no volatile loss; for Sol-BSP, this amounts of 1.33-1.45 wt. % $Na_2O$. In contrast, BSE has just 0.13 – 0.36 wt. % $Na_2O$ (Fig. 2b). Mars, also shown, has <40% of the Sol-BSP value (Taylor 2013). Since < 0.5% of exoplanets have >3.55 wt. % $Na_2O$ (Fig. 2c), the vast majority of exoplanets (99.5%) will have <0.9 wt. % $Na_2O$ in their mantles if they retain Earth-like proportions of $Na_2O$, and <1.3 wt. % $Na_2O$ if they retain Mars-like proportions. Most of this Na can likely be absorbed by pyroxene and garnet. For example, mean MORB has 2.79 wt. % $Na_2O$, and at high pressures, this Na can is absorbed by omphacite and garnet in eclogite. We thus suspect that minerals such as albite are rare in exoplanetary mantles.

## Carbon and Sulfur; probably unimportant for mantle mineralogy

In Fig. 2d, we calculate exoplanetary BSPs for all 2,543 stars in the Hypatia catalog where all our cations of interest (Na, Mg, Al, Si, Ca, Ti, Fe) and each of C and S are reported, with C treated as $CO_2$ (Supplementary Table 3). To these we compare estimates of Earth's mantle (McDonough and Sun 1995; Salters and Stracke 2003), Bulk Silicate Mars (Morgan and Anders 1979; Lodders and Fegley 1997), average CI chondrites (McDonough and Sun 1995), and Sol-BSP. All BSPs are calculated for the highly idealized case that all S and C are retained by a planet, and that neither enter the core. This assumption is, of course, unrealistic, but provides effective depletion factors due to a combination of nebular condensation, accretion, core formation and mantle devolatilization processes: in the events, Earth's mantle has retained <0.7% S and <0.006% of $CO_2$ relative to the idealized Sol-BSP case. This is not say that C and S contents are uninteresting. Carbon is clearly essential for life and its global cycle may affect climate (Sleep and Zahnle 2001). But absent new high *P-T* experiments that indicate the existence of refractory S- and C-rich phases, neither element appears crucial for understanding the dynamics of planetary interiors.

## Earth is Non-Solar (and Non-Chondritic)

We also find that Sol-BSP does not match published estimates of BSE. This means either (a) that published estimates of Bulk Silicate Earth (BSE) are in error or (b), a fundamental assumption in exoplanet composition studies is wrong: exoplanets do not precisely mimic the stars they orbit. Nickel provides a minor but perhaps significant example: a Solar bulk Earth (2.54 wt. % Ni) and a mantle with 0.196 wt. % Ni )McDonough and Sun 1995) yields $\alpha_{Ni}^{Earth} = 0.08$, but this requires a core with 7.3% Ni, greater than the 5% Ni usually inferred (e.g., Wood et al. 2014). Perhaps terrestrial Ni contents are uncertain, but Fe is more problematic. By design, applying $\alpha_{Fe}^{Earth} = 0.263$ yields a Sol-BSP with a BSE-like FeOt of 8 wt. %, and this yields a pure-Fe core radius, $r_{Sol-BSP}^{core}$ of 3,483 km, assuming mean core density is 11,247 kg/km$^3$ (from the Preliminary Reference Earth Model or PREM; Anderson 1989)—stunningly and deceivingly close to Earth's 3,480 km. But SiO$_2$ for the resulting Sol-BSP is high, at 48.4% (or 48.9% if we add no Si to the core), compared to 45% for BSE—a difference that is not trivial from a mineralogical perspective. And of course, Earth's core is not pure Fe but as low as 85% Fe (Rubie et al. 2011; Wood et al. 2014). If we maintain $\alpha_{Fe}^{Earth} = 0.263$, no reasonable amount of Si in the core brings SiO$_2$ contents into agreement. If we instead obtain Fe$^{BSP}$ by subtracting Earth's core (assuming it is 87% Fe), then $\alpha_{Fe}^{Earth} = 0.494$, and Sol-BSP has a BSE-like SiO$_2$ (44.9 wt. %), but much greater FeOt (15.6 wt. %; Fig. 2a) and lower MgO (30.8% vs. BSE's ca. 38-39%; Table 1). To yield a Sol-BSP with 8 wt. % FeOt at $\alpha_{Fe}^{Earth} = 0.494$, Earth's core would have to be >40% of Earth's mass, instead of the observed 33%. In summary, no value of $\alpha_{Fe}^{Earth}$ yields a Sol-BSP that simultaneously matches BSE's SiO$_2$, MgO, and FeOt (Fig. 2a); and hence the trend of Sol-BSP estimates obtained using different $\alpha_{Fe}^{Earth}$ values do not intersect BSEs in Fig. 2.

## Exoplanet Mineralogy and Temperature-derived Uncertainty

Figure 3 shows that while most exoplanets plot as peridotites or pyroxenites, the dominant variation in exoplanet mineralogy involves a tradeoff between Ol and Opx; variations in Cpx (and garnet, not shown) are small in comparison, because across the Hypatia Catalog, the ranges of CaO and Al$_2$O$_3$ are small compared to ranges in SiO$_2$, FeO and MgO. Figure 3 also shows how mineralogy varies as a function of $\alpha_{Fe}$. We anticipate that exoplanetary systems will likely have planets that exhibit a range of mineralogies (even for a given Mg/Si), just as $\alpha_{Fe}$ varies within our solar system. To illustrate the magnitude of the effect of $\alpha_{Fe}$, we plot median exoplanet mineralogies as large black crosses in Fig. 3: for $\alpha_{Fe} = 0.263$, median exoplanets have 38% Ol. But for a Mars-like $\alpha_{Fe}$ 0.537, median Ol rises to 60%, and for a Mercury-like $\alpha_{Fe} = 0.0$, mean Ol drops to 19%. This shift is directly related to how Fe is partitioned between core and mantle: adding Fe from the core to the mantle (lower $\alpha_{Fe}$) increases (FeO+MgO)/SiO$_2$ and effectively dilutes Si in the mantle, which increases Ol content; in contrast, placing more Fe into the metallic core reduces (FeO+MgO)/SiO$_2$ and SiO$_2$ is enriched in the residual mantle, which increases Opx. We also find that an assumed planetary temperature (*T*) also affects mineral proportions (Fig. 3). All of the plotted compositions in Fig. 3 assume a high-*T* case (1350°C; see Methods) where minerals exhibit extensive solid solution. At low *T* (<950°C), solid solution would decrease; the black arrows in Fig. 3c illustrate the strong shift towards Ol with lesser degrees of solid solution. To understand why, consider that Fig. 3 effectively illustrates a competition between MgO- (Ol), SiO$_2$- (Opx) and CaO-rich (Cpx) phases. At low *T*, Opx dissolves less Al and more Si into its tetrahedral site (as does Cpx), and so at low-*T*, less Opx is needed to accommodate the same amount of SiO$_2$ in a given bulk composition, so a system shifts towards the Ol apex. With smaller total amounts of SiO$_2$ in a bulk composition (compositions that are already near the Ol apex) the magnitude of the shift is less. Shifts toward or away from Cpx are subdued and affected by renormalization in the projection from garnet; this is because total Cpx is controlled by Ca, which is a minor component in planetary bulk compositions; in isolation, higher Ca (i.e., lower *T*) in Cpx shifts mineral proportions away from the Cpx apex. Interstellar variations in Ca and Al provide minor shifts in normative Cpx and Gar abundances, but these elements are too low in total abundance to require other Ca- or Al-rich phases.

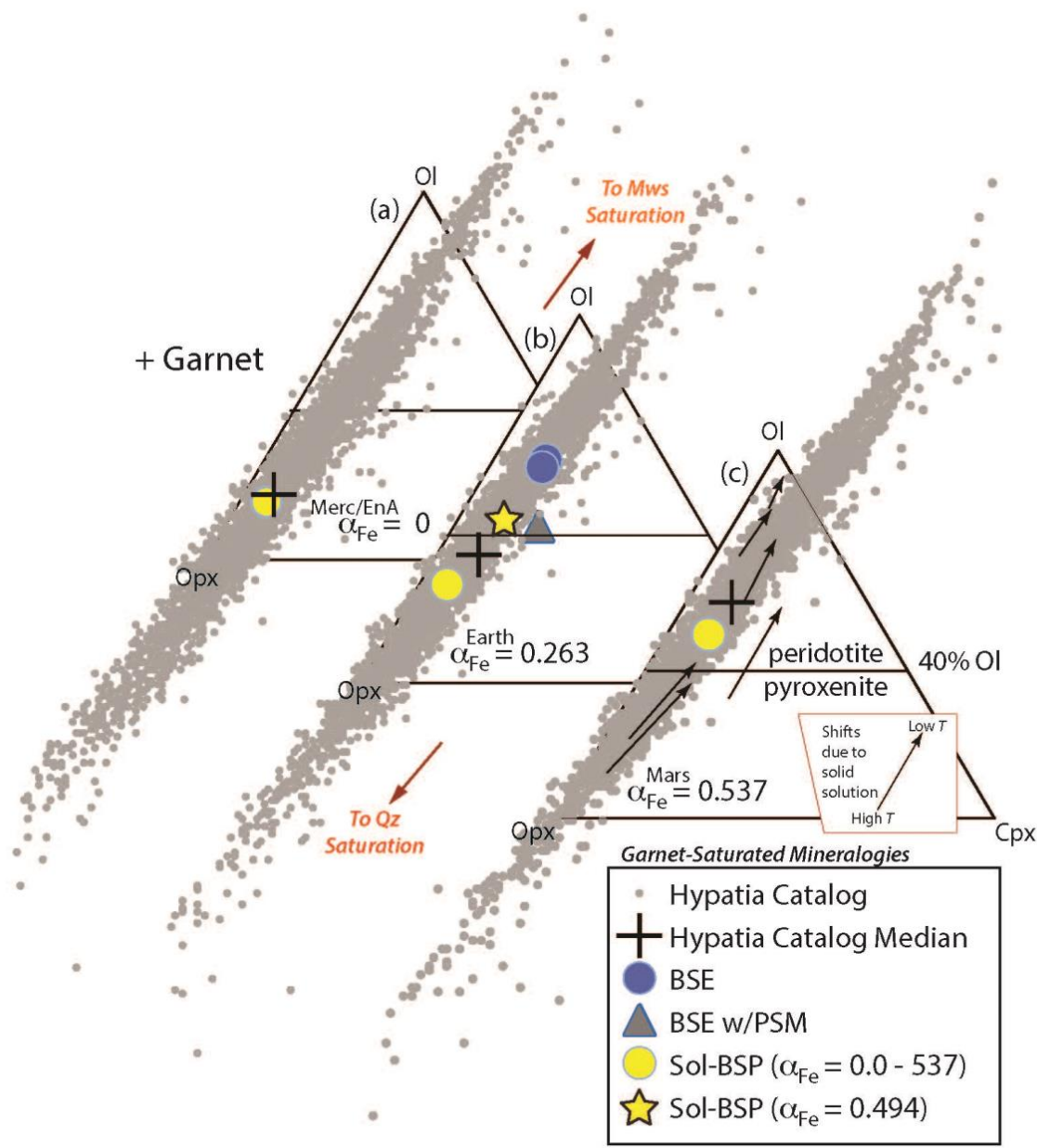

**Figure 3.** Exoplanet BSPs (EBSP) are plotted using the ultramafic rock diagram of Le Bas and Streckeisen (1991) Major oxides of EBSPs (see Appendix A) are recast as the mineral components: Olivine (Ol) = $Fm_2SiO_4$, Orthopyroxene (Opx) = $Fm_{1.9}Ca_{0.1}Al_{0.2}Si_{1.8}O_6$, Clinopyroxene (Cpx) = $Ca_{0.6}Fm_{1.4}Al_{0.2}Si_{1.8}O_6$ and Garnet = $Fm_{2.7}Ca_{0.3}Al_2Si_3O_{12}$, where FmO = FeO+MgO; the Ol-Opx-Cpx proportions are projected from garnet; mineral compositions approximate Earth's current upper mantle, at 1350°C and reproduce mineral proportions of Workman and Hart (2005) to within 10%. Mantle mineralogy is highly sensitive to how Fe is partitioned between a silicate mantle and a metallic core, as illustrated for (a) a possibly Mercury-like or Enstatite Achondrite (EnA) case, where $\alpha_{Fe}$= 0.0, (noted as "Merc-EnA"), (b) a solar bulk Earth-like scenario, with $\alpha_{Fe}$= 0.263 and (c) a Mars-like case, with $\alpha_{Fe}$= 0.537. Panel (b) also shows where Sol-BSP would plot if $\alpha_{Fe}$= 0.494. Other bulk compositions are as in Fig. 1. In (c) we also show black arrows to indicate how mineral proportions are affected by solid solution, as the plotted high *T* minerals are replaced by low *T*, pure end-member compositions, i.e, Olivine = $Fm_2SiO_4$; Orthopyroxene = $Fm_2Si_2O_6$; Clinopyroxene = $CaFmSi_2O_6$; Garnet = $Fm_3Al_2Si_3O_{12}$. The net effect is to yield higher contents of garnet and olivine, and lesser amounts of Opx and Cpx. Plotted positions, especially Cpx contents, are affected by renormalization in the projection from garnet.

Other Sources of Variability and Uncertainty

A source of uncertainty arises from the "spread" in published stellar compositions (Hinkel et al. 2014), which represent the range of reported values of elemental concentrations. When translated to mineral proportions, the magnitude of the "spread" translates to slightly less than the size of the large circles (for Sol-BSP and BSE) in Fig. 3. Current experiments, and our tests of mineral fraction estimates (Fig 1) use bulk compositions that mostly range of 40-49 wt. % $SiO_2$, and 20-50 wt. % MgO, but many EBSPs fall outside these ranges; new experiments may reveal unique mineral assemblages that are not extant among terrestrial mantle samples.

# Discussion

An Exoplanet Classification Model & Sources of Uncertainty

However imperfect our mineral estimation methods may be, we can be certain from the Hypatia Catalog (Hinkel et al. 2014) that many if not most exoplanets will have silicate mantles that mimic lithologies found on Earth (Figs. 2a, b), and that either or both of Ol and Opx will be dominant (Fig. 3). Some planets may approach monomineralic dunite or orthopyroxenite, and some have sufficient MgO so as to be magnesiowüstite (Msw)-saturated (Fig. 2b), or sufficient $SiO_2$ to be saturated in quartz (Table 2). But no exoplanets appear wholly exotic (i.e., have mantles made of rutile or corundum or albite, etc.) and even variations in Cpx (Fig. 3) or garnet are narrow.

Figure 4 illustrates our proposed classification scheme based on the ratio (FeO+MgO)/$SiO_2$ for a given Bulk Silicate Planet (BSP) composition (Table 2; Fig. 4). This approach does not yield mineral proportions, but separates observable rock types (peridotites and pyroxenites) and solid solution limits of olivine and orthopyroxene. Such estimates of mantle rock type, however, are very sensitive to assumed values of $\alpha_{Fe}$: as $\alpha_{Fe}$, varies from 0 to 0.537 (Fig. 3) mean exoplanetary Ol contents shift from about 20% to 60% Ol (Fig. 3), and mantle rock types shift from 59% pyroxenite to 95% peridotite (Table 2). In addition, planets do not necessarily mimic the stars they orbit; Mars is plausibly solar in bulk composition (Figs. 4d-f) but Earth is non-solar (Figs. 2-3; Figs. 4a-c ). And predicting how exoplanets might deviate from a stellar composition might not be straightforward. For example, Palme (2000) suggests that no compositional gradients exist from Mercury to Vesta, and we find the same: $\alpha_{Fe}$ increases from near 0 to 0.54 from Mercury to Mars, but $\alpha_{Fe}$ = 0.13 at Vesta, using Steenstra et al. (2016). So planet-star distance might not be helpful in narrowing an exoplanetary bulk composition.

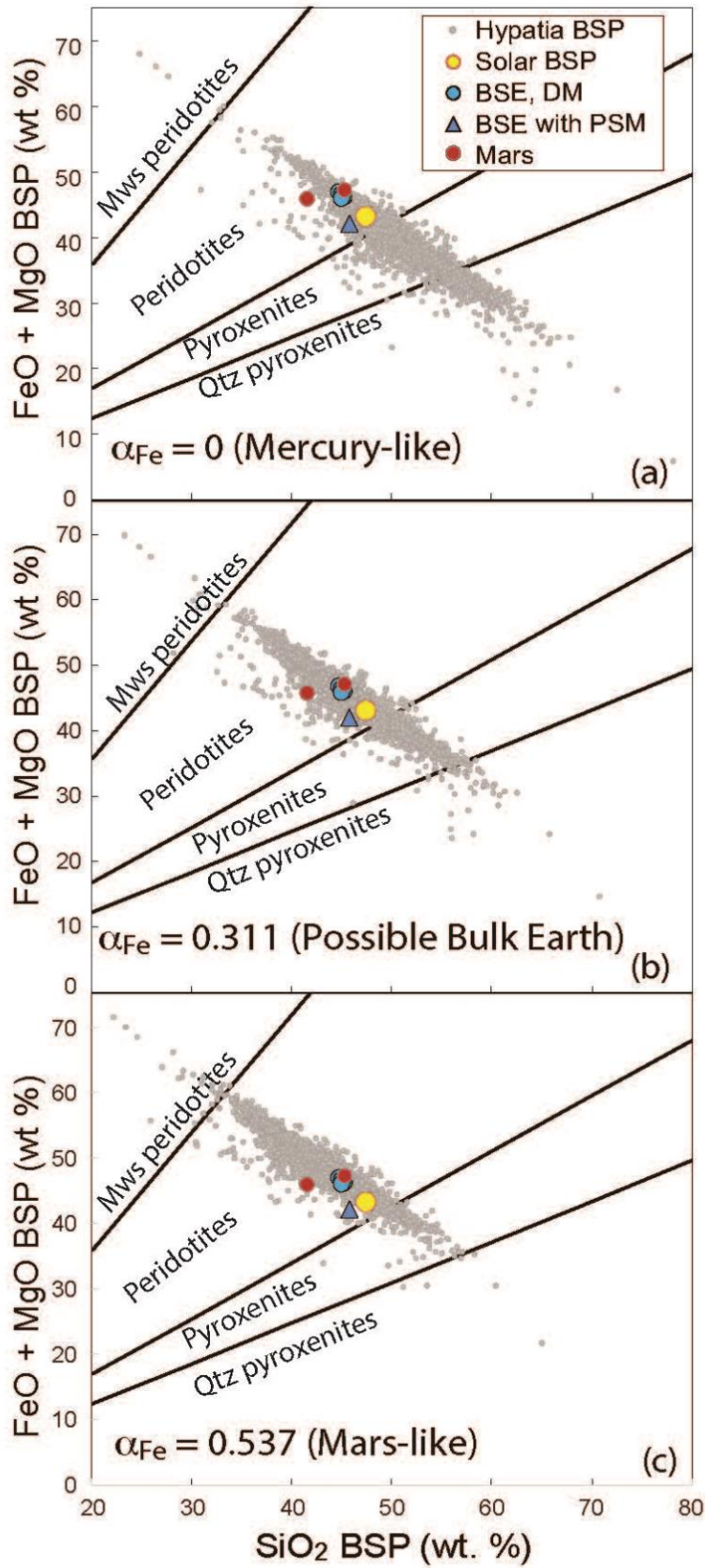

**Figure 4.** Our proposed rock classification scheme for exoplanets, based on the ratio [FeO+MgO]/SiO$_2$ (Table 2), for the cases of (a) $\alpha_{Fe}$ = 0 (Mercury- or Enstatite Achonrdite Parent Body-like), (b) $\alpha_{Fe}$ = 0.263 (Solar) and (c) $\alpha_{Fe}$ = 0.54 (Mars-like). Mws = magensiowüstite; Qtz = quartz. Other abbreviations, data sources and calculations are as in Fig. 1. The indicated rock types (Mws-normative peridotite, peridotite, pyroxenite, Qtz-normative pyroxenite) are based on our normative or standard mineral compositions, noted in Fig. 3 caption. If end-member mineral compositions are used instead, BSPs shift towards Mws-normative peridotite. Note that contrasts in the partitioning of Fe significantly affect bulk silicate planet classification.

## Peridotite vs. Pyroxenite? The Differences Might be Important

Seemingly subtle differences in mantle mineralogy may be important. Compared to a peridotite mantle, a pyroxenite-rich exoplanet may yield thicker crust (Lambart et al. 2016), albeit still basaltic in composition (Lambart et al. 2009). It is not clear what affect this might have on plate tectonics. A thick, rigid, Opx-rich crust, might be stronger (e.g, Yamamoto et al. 2008), and less likely to break into plates. Alternatively, with sufficient water, Opx-rich systems might partially distill into very thick silicic crustal bodies, and so enhance crustal and lithosphere density contrasts. New experiments are needed on the melting and yield strength behaviors of MgO- and Opx-rich compositions (Supplementary Tables 2-3) to understand these systems.

## Earth is Non-Solar/Non-Chondritic

We verify a long-standing result: Earth's bulk composition is non-solar and non-chondritic (e.g., McDonough and Sun 1995; Drake and Righter, 2002). The problem of a Solar bulk Earth with respect to the non-volatile elements is encapsulated by the tradeoff in predicting $SiO_2$ or FeO: if $\alpha_{Fe}^{Earth} = 0.263$ (a solar bulk composition and a BSE with 8 wt. % FeO), then Earth's mantle has less $SiO_2$ than carbonaceous chondrites which, as expected, are similar to Sol-BSP (Figs. 2a, 5b). And the mis-match in $SiO_2$ is worse still for enstatite chondrites (Fig. 5a-c). But if we apply $\alpha_{Fe} = 0.494$ (a solar bulk Earth with a core having 87% Fe and unknown mantle FeO content), this yields a BSE-like $SiO_2$ (Fig. 2a, 5e), but FeO contents much higher than anything presumed for BSE (Figs. 2a, 5e). If we instead allow that Si has been lost to an early atmosphere (Fegley et al. 2016) or to the core (Wade and Wood 2005), Earth still has higher-than-solar CaO and $Al_2O_3$ (Fig. 5c). The contrasts in $SiO_2$ may seem minor, but they make a tremendous difference in mantle mineralogy: at $\alpha_{Fe}^{Earth} = 0.263$, Sol-BSP plots solidly in the pyroxenite field and at $\alpha_{Fe}^{Earth} = 0.494$ it is still Ol-poor compared to BSE (Fig. 3). These results might appear to bolster arguments for a pyroxenite-rich mantle, derived by adding subducted mid-ocean ridge basalt (MORB) to depleted mantle (e.g., Hirschmann and Stolper 1996; Sobolev et al. 2007; van der Hilst and Karason 2000). But if the true BSE is obtained by adding MORB, then BSE moves further from Sol-BSP with respect to FeOt (Fig. 2a), CaO and $Al_2O_3$ (Fig. 5c).

**Figure 5 (below).** In (a), (b) and (c) we compare estimates of depleted mantle (DM) and bulk silicate Earth (BSE) from Fig. 1, and bulk silicate lunar compositions from Kahn et al. (2006), to BSP estimates using carbonaceous (CC) and enstatite (EC) chondrite bulk compositions from Nittler et al. (2004) as starting compositions, assuming $\alpha_{Fe} = 0.263$. Also shown are bridgmanite (Mg-perovskite, Mg-pv) and majorite compositions from Hirose (2002). In (d), (e) and (f), we compare DM, BSE and chondrite-derived BSPs using a Mars-like $\alpha_{Fe}= 0.537$, with estimates of martian bulk silicate compositions as in Lodders and Fegley (1997). Panels (a)-(c) illustrate the challenge in using the Sun or chondrites as a bulk composition for Earth: while BSPs using enstatite chondrites (ECs) have CaO and $Al_2O_3$ contents that are too low, the most CaO- and $Al_2O_3$-rich BSPs from carbonaceous chondrites (CCs) overlap with Earth. But the CC-derived BSPs have $SiO_2$ and MgO that are too low. Estimates for BSE can have lower MgO by adding MORB, but this drives BSE estimates away from CCs with respect to CaO- and $Al_2O_3$. Hiding large amounts of bridgmanite and majorite might yield a Solar bulk Earth but only very specific bridgmanite compositions allow BSE to have lower $SiO_2$ (a) and near-Solar FeO (b).

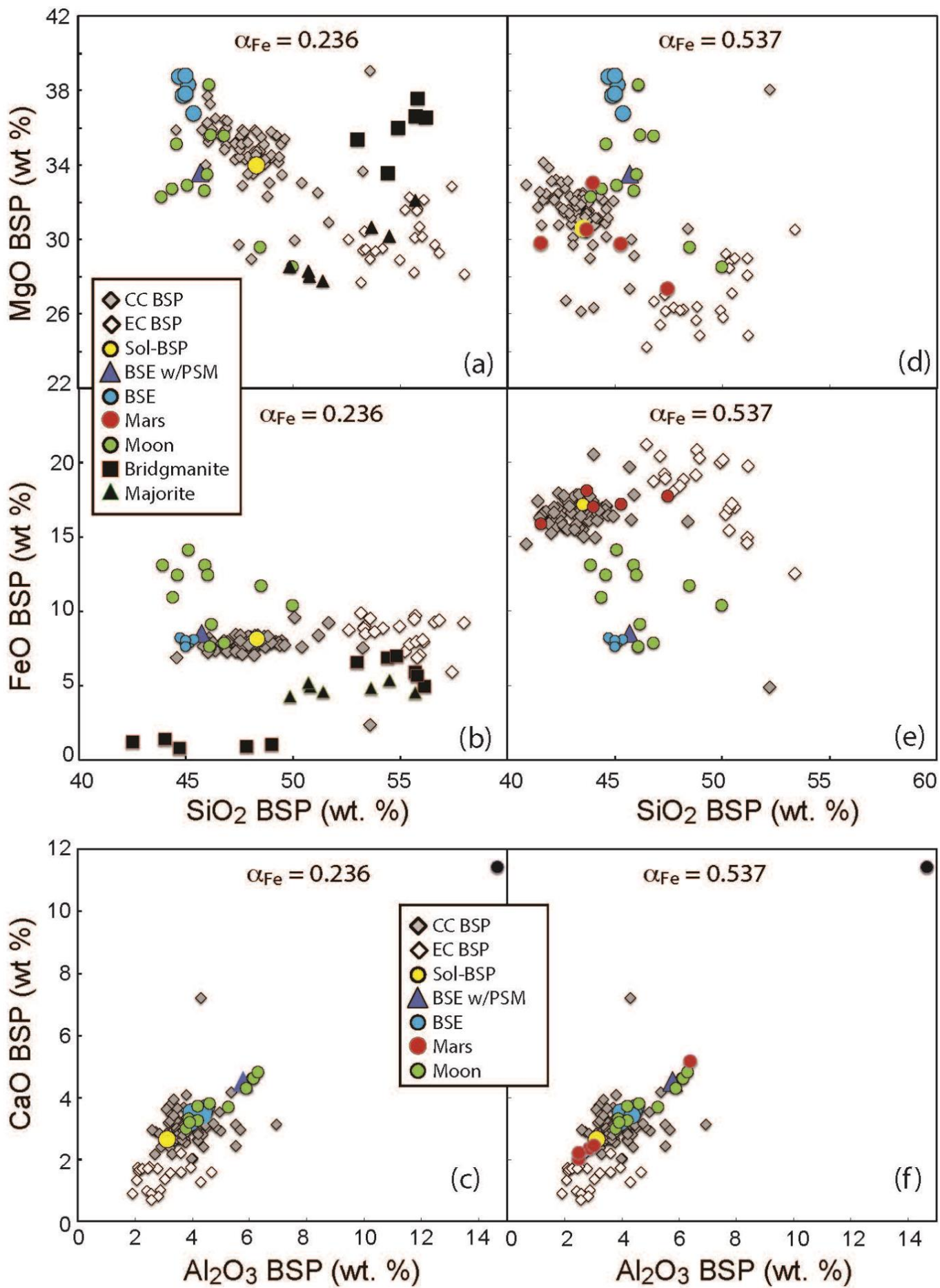

Does Earth Have a Hidden Mantle Component or a Compositionally Distinct Lower Mantle?

A remaining and alluring alternative to obtain a solar or chondritic bulk Earth is to assume that Earth's mantle contains a hidden component, $h$. Seismological arguments have long been advanced to yield a lower mantle that is enriched in FeO (e.g., Anderson and Jordan 1970; van der Hilst and Karason 2000), or $SiO_2$ (e.g., Murakami et al. 2012), although these perennially fail re-inspection (e.g., Davies, 1974; Irfune et al. 2010; Davies et al. 2012; Hyung et al. 2016). In any case, Agee and Walker (1988) posit a viable mechanism: high-pressure phases, such as bridgmantie or majorite (both nominally $(Mg,Fe)SiO_3$ but dissolving variable Ca, Al and other cations) might crystallize from an ancient magma ocean and settle into the lower mantle. The possibility of a hidden component, un-excavated by mantle plumes, has also gained support from isotopic studies that show Earth as non-chondritic, but where it assumed to be so (Boyet and Carlson 2005; Bouvier and Boyet 2016). Figure 5 shows that adding near-equal amounts of bridgmanite and majorite to BSE could yield Solar-like MgO and $SiO_2$ (Fig. 5a); such additions might yield too low a value of FeO (Fig. 5b), but the mineralogy of the deep mantle is not well known and other phases might compensate, so the idea of Agee and Walker (1988) is possible.

As a further test, we delimit the major oxide compositional range of $h$ (Table 2) by subtracting from a solar bulk composition (a) Earth's core (assuming 87 wt. % Fe), and (b) the depleted mantle (DM) component that is needed to feed mid-ocean ridge basalts. In this model, the fraction of the hidden component, $F_h$ and the fraction of DM, $F_{DM}$ are both unknown, but they sum to unity (note that continental crust is too low in abundance to affect major oxide mass balance), and the sum is equivalent to bulk silicate Earth at $\alpha_{Fe} = 0.494$: so $F_h + F_{DM} = 1 = $ Sol-BSP($\alpha_{Fe} = 0.494$). This mass balance test shows that $h$ must be >28% of Earth's mantle, otherwise $Al_2O_3$ and $TiO_2$ in $h$ are negative (Table 1); this fraction is much greater than the <3% volume estimated for seismic anomalies hypothesized to represent a hidden component (see He and Wen 2012). It would represent everything below about 1,850 km, and so approximate the bottom 1,000 km of the mantle (e.g., Albarede and van der Hilst 2002), but this component would be effectively Ca- and Al-free. For the case where $F_h = 0.73$ (where $F_h$ is all of Earth's lower mantle) no composition of $h$ yields a plume source mantle (PSM; Table 1) that would feed ocean island basalts (Table 1), as FeO contents are too high and CaO and $Al_2O_3$ contents are too low (Table 1). If we adopt Javoy et al.'s (2010) enstatite chondrite bulk Earth (Javoy et al. 2010) we obtain a lower mantle that is much less extreme in its Fe-enrichment (9.24 wt. % FeO), but $SiO_2$ is quite high (51.5 wt. %), and MgO is low (35.2 wt. %), as are CaO (1.4 wt. %) and $Al_2O_3$ (1.8 wt. %), especially relative to PSM.

In summary, if $h$ exists, it is not only quite large, but it is *not* the source of mantle plumes (PSM in Table 2) and it is enriched in $Na_2O$ and FeO, and depleted in MgO, CaO and $Al_2O_3$, compared to any estimates of BSE or DM. Our conclusion is to reject $h$ since to accept it requires rejecting long-standing arguments in support of whole-mantle convection (e.g., Davies, 1977), the nucleation of thermal plumes near the core/mantle boundary (Davies 1988; Davies and Richards 1992), and thermal models (Farnetani 1997) that explain observed plume excess temperatures (Putirka 2005; Putirka et al. 2007). Especially compelling are seismic images of subducted slabs that reach the base of the mantle (Jordan 1977; Grand 1997; Fukao and Obayashi 2013); these require a return flow, and appear to end discussion of an isolated lower mantle. Numerical models also appear to resolve supposed conflicts between whole-mantle convection and a mantle with geochemical and seismic heterogeneity (Jordan et al. 1993; Schuberth et al. 2009; Barry et al. 2018). So while controversy still exists over seismic images of deep-seated plumes (e.g., Montelli et al. 2004), we agree with recent Agrusta et al. (2018), that Earth's mantle has long been well stirred. But if all these studies are in error, Table 2 provides estimates for lower mantle composition, for three cases: that $h$ exists below 1,850 km (a minimum volume), 1,000 km and 660 km.

# Implications

Although exoplanet compositions can vary widely between pyroxenite and peridotite, even within a given exoplanetary system, our results indicate no obvious barrier to the evolution of exoplanetary felsic (continental) crust, or plate tectonics. For example, metallic cores may be proportionately smaller or larger than Earth's, but for Earth-sized planets, no mantle depths would be too shallow to limit mantle convection. Bystricky et a. (2016) suggest that the viscosity of Opx might not be much different than Ol at high temperatures, and so our observed mineralogical contrasts should not greatly affect mantle viscosity and convection.

Our estimates of BSPs should also be useful where bulk densities of specific exoplanets are being measured (e.g., Gillon et al. 2017); if core size is known, then mantle mineralogy can be more precisely estimated, and it might then be possible to apply studies such as Weller and Lenardic (2018) to evaluate tectonic potential of specific planetary bodies. In their Fig. 5b, Weller and Lenardic (2018) compare temperature and yield strength to discriminate between plate tectonic and stagnant lid convective regimes. Our results show that the key issue is whether yield strength is affected by the mantle Ol/Opx ratios (Fig. 4). Current experiments leave the issue clouded. Yamamoto et al. (2008) and Bystricky et al. (2016) both propose that the yield strength for Opx is greater than for Ol at least at low temperatures, but Hansen and Warren (2015) suggest that differences in viscosity between Opx and Ol are small. Large contrasts in yield strength at shallow depths (low temperatures) might mean that Opx-dominated exoplanets have strong lithospheres that resist tectonic deformation. But Opx-rich exoplanets might also yield thicker and more abundant basaltic crust (Lambart et al. 2009, 2016), which might then distill into $SiO_2$-rich (granitic) crust (e.g., Sisson et al. 2005), which could heighten lithospheric density contrasts and enhance tectonic initiation. These speculations require experimental numerical tests.

Finally, efforts to use chondrites as a possible bulk composition for Earth (e.g., Boyet and Carlson 2005; Fitoussi et al. 2009; Javoy et al. 2010) appear misguided. We agree with McDonough and Sun (1995) that chondrite meteorites "are not the main building blocks of Earth". And our major oxide comparisons (Fig. 5) support the specific conclusion of Drake and Righter (2005), that our planet is constructed of "Earth Chondrites". Estimates of Bulk Silicate Earth $SiO_2$-MgO-FeO plot at the edge of or outside the array of carbonaceous and enstatite chondrite meteorites (Fig. 5). The lack of chondrites in the vicinity of BSE may mean that the planetesimals that formed Earth were consumed during Earth's accretion (e.g., the "Earth Chondrites" of Drake and Righter 2005). This does not mean that planets never match the composition of the stars they orbit. Mars is plausibly solar (or carbonaceous chondritic) in bulk composition (Figs. 5d-f). But perhaps chondrites have always been a red herring. Hewins and Herzberg (1996) proposed that chondrules might be an important component in Earth. Iron contents seem too low for chondrules to comprise bulk Earth, but Connolly and Jones (2016) suggest that they are among the most abundant components of the early Solar System. To better understand exoplanets, we need to dispel the shadow that is cast over knowledge of our own early Solar System.

**Acknowledgments** We appreciate reviews by Anne Hofmeister, Natalie Hinkel, and Rhian Jones. Comments by Hinkel and Jones especially, led to many clarifying edits to the paper and a re-thinking of key assumptions. Some very insightful comments by Jones also forced us to replot most of our data, and correct a crucial error with regard to our meteorite comparisons. We are also immensely grateful for access to the Hypatia Catalog Database, an online compilation of stellar abundance data as described in [Hinkel et al. (2014, AJ, 148, 54)](#), which was supported by NASA's Nexus for Exoplanet System Science (NExSS) research coordination network and the Vanderbilt Initiative in Data-Intensive Astrophysics (VIDA). Alain Plattner and John Wakabayashi also provided comments on early drafts of this and a related manuscript; Plattner's questions prompted our investigation of $a_{Fe}$. We especially thank Natalie Hinkel for answering our many questions about the Hypatia Catalog, to Cayman Unterborn for sharing a version of his spreadsheet for transforming stellar compositions from dex to major oxides, and to both Natalie and Cayman for sharing pre-prints of their work, and explaining ideas and methods expressed in these.

Submitted to *American Mineralogist*formation in Vesta from metal-silicate partitioning of siderophile elements. Geochimica et Cosmochimica Acta 177, 48-61.

Stern, R.J. (2006) Is plate tectonics needed to evolve technological species on exoplanets? Geoscience Frontiers, 7 573-580.

Stevenson, D.J. (2003) Styles of mantle convection and their influence on planetary evolution. Geodynamics, 335, 99-111.

Tachinami C, Senshu H, and Ida, S. (2011) Thermal evolution and lifetime of intrinsic magnetic fields of super-earths in habitable zones. Astrophysical Journal, 726, doi.org/10.1088/0004-637X/726/2/70.

Taylor, G.J. (2013) The bulk composition of Mars. Chemie der Erde, 73, 401-420.

Thiabaud, A., Marboeuf, U., Alibert, Y., Leya, I., and Mezger, K. (2015a) Elemental ratios in stars vs. planets. Astronomy and Astrophysics, 580, A30, doi: 10.1051/0004-6361/201525963.

Thiabaud, A., Marboeuf, U., Alibert, Y., Leya, I., and Mezger, K. (2015b) Gas composition of the main volatile elements in protoplanetary discs and its implication for planet formation. Astronomy and Astrophysics, 574, A138, doi: 10.1051/0004-6361/201424868.

Thompson, J.B. (1982) Reaction Space: An algebraic and geometric approach. In, Ferry, J.M. ed., Characterization of Metamorphism Through Mineral Equilibria. Reviews in Mineralogy, 10, 33-52, Mineralogical Society of America, Washington D.C.

Thompson, S.E. et al. (2018) Planetary candidates observed by Kepler. VIII. A fully automated catalog with measured completeness and reliability based on Data Release 25. Astrophysical Journal Supplement, 235, doi:10.3847/1538-4365/aab4f9.

Unterborn, C.T., Hull, S.D., Stixrude, L., Teske, J.K., Johnson, J.A., and Panero, W.R. (2017) Stellar chemical clues as to the rarity of exoplanetary tectonics. arXiv:1706.10282v2 [astro-ph.EP] 3 Jul 2017.

Unterborn, C.T. and Panero, W.R. (2017) The effects of Mg/Si on the exoplanetary refractory oxygen budget. The Astrophysical Journal, 845:61 (9pp).

Vanderspek et al. (2018) TESS discovery of an ultra-short-period planet around the nearby M dwarf LHS 3844. arXiv:1809.07242v1.

van der Hilst, R.D., and Karason, H. (2000) Compositional heterogeneity in the bottom 1000 kilometers of Earth's mantle: toward a hybrid convection model. Science, 283, 1885-1888.

Wade, J. and Wood, B.J. (2005) Core formation and the oxidation state of the Earth. Earth and Planetary Science Letters, 236, 78-95.

Walter, M.J. (1998) Melting of garnet peridotite and the origin of komatiite and depleted lithosphere. Journal of Petrology, 39, 29-60.

Warren, J.M. (2016) Global varaitions in abyssal peridotite compositions. Lithos, 248-251, 193-219.

Weller, M.B., and Lenardic, A. (2018) O the evolution of terrestrial planets: bi-stability, stochastic effects, and the non-uniqueness of tectonic states. Geoscience Frontiers, 9, 91-201.

Wood, B.J., Kiseeva, E. and Mirolo, F.J. (2014) Accretion and core formation: the effects of sulfur on metal-silicate partition coefficients. Geochimica et Cosmocimica Acta, 145, 248-267.

Workman, R.K. and Hart, S.R. (2005) Major and trace element composition of the depleted MORB mantle (DMM). Earth and Planetary Science Letters, 231, 53-72.
21

**Table 1. Bulk Compositions**

| | Sol-BSP[1a] | Sol-BSP[1b] | Sol-BSP[1c] | Sol | Continental Crust[2] | MORB[3] | DM[4] | BSE-M&S[5a] | BSE-M&S[5b] | PSM[6] | BSE (if PSM = LM)[7] | Hidden[8a] $F_h$=0.28 | Hidden[8b] $F_h$=0.59 | Hidden[8c] $F_h$=0.73 |
|---|---|---|---|---|---|---|---|---|---|---|---|---|---|---|
| $SiO_2$ | 48.9 | 44.9 | 48.1 | 33.9 | 60.6 | 50.5 | 44.7 | 45.0 | 49.9 | 46.0 | 45.7 | 42.8 | 43.9 | 44.1 |
| $TiO_2$ | 0.16 | 0.14 | 0.15 | 0.11 | 0.70 | 1.68 | 0.13 | 0.20 | 0.16 | 0.47 | 0.38 | 0.00 | 0.11 | 0.12 |
| $Al_2O_3$ | 3.45 | 3.2 | 3.4 | 2.4 | 15.9 | 14.7 | 4.0 | 4.45 | 3.65 | 6.3 | 5.7 | 0.01 | 2.35 | 2.74 |
| $Cr_2O_3$ | 0.80 | 0.73 | 0.78 | 0.55 | - | 0.07 | 0.57 | 0.38 | 0.44 | 0.46 | 0.49 | 1.66 | 0.99 | 0.87 |
| FeOt | 8.07 | 15.6 | 9.60 | 34.5 | 6.70 | 10.43 | 8.18 | 8.05 | 8.0 | 8.68 | 8.54 | 35.3 | 20.9 | 18.5 |
| MnO | 0.52 | 0.48 | 0.52 | 0.36 | 0.10 | 0.18 | 0.13 | 0.14 | 0.13 | 0.14 | 0.14 | 1.41 | 0.73 | 0.62 |
| MgO | 33.6 | 30.8 | 33.0 | 23.2 | 4.7 | 7.6 | 38.7 | 37.8 | 35.2 | 31.9 | 33.6 | 14.3 | 26.6 | 28.7 |
| CaO | 2.9 | 2.6 | 2.83 | 1.99 | 6.4 | 11.4 | 3.2 | 3.6 | 2.9 | 5.0 | 4.5 | 1.28 | 2.47 | 2.67 |
| $Na_2O$ | 1.45 | 1.33 | 1.43 | 1.00 | 3.10 | 2.79 | 0.13 | 0.36 | 0.34 | 0.72 | 0.57 | 3.90 | 2.03 | 1.71 |
| $K_2O$ | - | - | - | - | 1.80 | 0.16 | 0.01 | 0.03 | 0.02 | 0.04 | 0.04 | - | - | - |
| $P_2O_5$ | - | - | - | - | 0.13 | 0.18 | 0.02 | 0.02 | - | 0.06 | 0.05 | - | - | - |
| NiO | 0.14 | 0.15 | 0.14 | 2.01 | - | 0.01 | 0.25 | - | - | 0.20 | 0.21 | 0.30 | 0.27 | 0.26 |

[1a]Bulk Silicate Planet obtained using $Fe^{BSP}/Fe^{BP} = \alpha = 0.263$, from Lodders (2010) solar composition and Earth's mantle FeO (8 wt. %); [1b]Sol-BSP obtained using $\alpha = 0.494$, from Lodders (2010) solar composition minus Fe from the core (87 wt. % Fe, 33% Earth's mass); [1c] $\alpha = 0.311$ uses Bulk Silicate Earth (McDonough and Sun 1995) and Earth's metallic core to obtain bulk Earth; [1d]Lodders (2010) solar composition as major oxides; [2]Rudnick and Gao (2003), average continental crust. [3]Gale et al. (2013), mean MORB; [4]Workman and Hart (2005), Depleted MORB Mantle; [5a]McDonough and Sun (1995) Silicate Earth, Pyrolite Mantle 1 and [5b]CI [meteorite] model; [6]PSM = Plume Source Mantle, this study; [7]Bulk Silicate Earth (BSE), if PSM if the Lower Mantle (LM) = PSM, i.e., PSM = mantle composition between 660 km and 2890 km. [8a]Hidden mantle component if Earth (with a 33% core mass, of 87% Fe, 5% Ni and 7% Si) has solar relative abundances of Si, Mg, and Fe, etc., and the hidden components is 28% of Earth's mantle (a minimum, otherwise Al and Ti are negative), which is equivalent to depths >1,850 km, or about the bottom 1000 km of the mantle; [8c]hidden component comprises all of the mantle below 1000 km. [8c]hidden component when it comprises all the lower mantle (below 660 km).

**Table 2. Classification of Exoplanetary Mantle Compositions**

| Exoplanet Mantle Composition (wt. % oxides)[1] | Standard Mineralogy | % Exoplanets $\alpha_{Fe} = 0.0$ | % Exoplanets $\alpha_{Fe} = 0.311$ | % Exoplanets $\alpha_{Fe} = 0.537$ |
|---|---|---|---|---|
| $\frac{MgO + FeOt}{SiO_2} < 0.62$ | Quartz-normative pyroxenite | 13.3 | 1.2 | 0.2 |
| $0.62 < \frac{MgO + FeOt}{SiO_2} < 0.85$ | Pyroxenite | 59.8 | 20.6 | 4.2 |
| $0.85 < \frac{MgO + FeOt}{SiO_2} < 1.8$ | Peridotite | 26.8 | 78.0 | 95.1 |
| $\frac{MgO + FeOt}{SiO_2} > 1.8$ | Magnesiowüstite-normative Peridotite | 0.1 | 0.2 | 0.5 |

[1]Boundaries are drawn using limits of natural, terrestrial pyroxenite and peridotite compositions. We assume that Magnesiowüstite (Msw) saturation occurs near the upper limit of $(FeO+MgO)/SiO_2$ in Ol, and that quartz (Qtz) saturation occurs below the minimum $(FeO+MgO)/SiO_2$ of orthopyroxene. We would expect dunite mineralogies when $(FeO+MgO)/SiO_2$ is close to 1.8 and orthopyroxenite when $(FeO+MgO)/SiO_2$ is close to 0.62.